\documentclass[pre,aps,floats,superscriptaddress,floatfix,twocolumn]{revtex4}
\usepackage{amssymb,amsmath}
\usepackage{amsmath,amssymb}
\usepackage{graphicx}
\usepackage{psfrag}
\usepackage{color}
\usepackage{soul}
\usepackage{dcolumn}
\usepackage{bm}
\usepackage[normalem]{ulem}

\def\beq{\begin{equation}}
\def\eeq{\end{equation}}
\def\bea{\begin{eqnarray}}
\def\eea{\end{eqnarray}}
\begin{document}
\title{
%Uniform and nonuniform steady states in a periodic asymmetric exclusion process with defects }
%Varieties of steady states in a periodic asymmetric exclusion process with defects\\
%Defect versus defect: Steady states in a periodic heterogeneous asymmetric exclusion process\\
%Defect versus defect: phases of single file marching in periodic landscapes with road blocks
Steady states in asymmetric exclusion processes with reactions: Fluctuations and non-mean field behaviour
}
\author{Rohn Chatterjee}\email{rohn.ch@gmail.com}
%\affiliation{VECC}
\affiliation{Barasat Government College,
10, KNC Road, Gupta Colony, Barasat, Kolkata 700124,
West Bengal, India}
\author{Atri Goswami}\email{goswami.atri@gmail.com}
\affiliation{Gurudas College, 1/1, Suren Sarkar Road, Jewish Graveyard,
Phool Bagan, Narkeldanga, Kolkata 700054, West Bengal, India}
\affiliation{Barasat Government College,
10, KNC Road, Gupta Colony, Barasat, Kolkata 700124,
West Bengal, India}
\author{Sudip Mukherjee}\email{sudip.bat@gmail.com,aca.sudip@gmail.com}
%\author{SM}
\affiliation{Barasat Government College,
10, KNC Road, Gupta Colony, Barasat, Kolkata 700124,
West Bengal, India}

\date{\today}

\begin{abstract}
 We study an totally asymmetric simple exclusion process (TASEP) in an open geometry. We couple TASEP with simple reactions in the form of production (birth) and degradation (death) due to over crowding taking place in the bulk. 
 We investigate the steady states, which are generically space-dependent. Phase coexistences are also possible for some choices of the model parameters. We study the system by mean-field theories (MFT) supplemented by extensive Monte-Carlo simulations (MCS). Intriguingly, unlike oridinary TASEP and other related models, configuration-dependent bulk reactions in a TASEP model can make nearest-neighbour correlations quantitatively important even in an open driven system where mean-field theory would ordinarily be expected to provide a useful description. We attribute this to the correlation effects ignored in MFT. We numerically quantify this correlation effect.
 
\end{abstract}

\maketitle 

\section{Introduction}

Driven diffusive systems~\cite{driv1,driv2,driv3,driv4}  exhibit remarkably complex behavior despite being governed by purely local dynamics, even in one-dimensional settings. A prototypical example is the totally asymmetric simple exclusion process (TASEP)~\cite{krug-prl,derrida1,derrida2,derrida3}, which is defined on a one-dimensional lattice with open boundaries. The dynamics are stochastic and obey the exclusion principle, allowing at most one particle to occupy a site at any given time. Particles are injected into the lattice at one boundary with a specified rate, move unidirectionally from site to site with unit hopping rate, and exit through the opposite boundary at another prescribed rate. Depending on the particle injection and extraction rates, the TASEP exhibits three distinct steady-state phases: a low-density (LD) phase, characterized by a bulk density below one-half; a high-density (HD) phase, where the bulk density exceeds one-half; and a maximal-current (MC) phase, in which the bulk density is exactly one-half.

TASEP was originally conceived as a minimal model to describe proteins synthesis in biological cells~\cite{macdonald}. Subsequently it was found to display boundary-induced phase transitions~\cite{krug-prl,derrida1,derrida2}.
Many extensions of the TASEP have been developed to better model real-world transport processes. These include studies on transport through network-like structures~\cite{andrea1,rakesh1,rakesh2} inspired by molecular motor movement in cells~\cite{howard}, microtubule length regulation by kinesin-8 motors~\cite{melbinger1,melbinger2}, and two-lane TASEP systems with particle switching that can produce phenomena such as jams and domain-wall delocalization~\cite{reichenbach1,reichenbach2,reichenbach3}. Researchers have also investigated the effects of obstacles and bottlenecks - both static and dynamic - which can hinder particle movement, similar to traffic congestion or protein-induced hindrance in cellular transport; see Refs.~\cite{helbing2001,kerner1998,treiber2013,dixit2008,telley2009}. Such spatial variations in hopping rates can lead to nonuniform steady-state density profiles~\cite{lebo,niladri-tasep,kolomeisky1998,qiu2007,sm-tasep-atri3,pal-point-defect,shaw2004,mustansir,hinsch,tirtha-niladri,parna-anjan,tirtha-qkpz,pierobon2006,klumpp2005,waclaw2019,jindal2020,sahoo2014,tirtha-qkpz,sm-tasep-atri1,sm-ab-tasep}. Effects of multiple lanes~\cite{sm-tasep-atri2,erwin-sep-tasep1,erwin-sep-tasep2}, weak local nonconserving dynamics~\cite{ef-lktasep-prl,ef-lktasep-pre,tirtha-lk1,tirtha-lk2} and finite resources ~\cite{reser1,reser3,reser2,astik-1tasep,astik-erwin,sourav1,sourav2,sourav3,arvind1,sourav5} are also known to give rise to novel steady states.

Reaction–diffusion models~\cite{schutz,marro,hinrich,uwe} describe systems with  particles that undergo diffusion and local reactions such as creation, annihilation, coagulation, or branching. These models provide a simple framework for studying nonequilibrium phenomena, including phase transitions, pattern formation, and absorbing-state dynamics. Due to their analytical tractability and rich collective behavior, they have been extensively used in statistical physics, population dynamics, chemical kinetics, and biological transport processes.

In this article, we study one-dimensional (1D) local reactions with open boundaries in the presence of a drift or unidirectional motion. To that end, we combine the ideas of an open TASEP together with the simplest one-component reaction model. Our studies can serve as a conceptual model for population dynamics in the presence of in and out migration and chemical reactions in a moving environment. We investigate the steady states of the model. We  find that the stationary densities in general have complex dependence on space (i.e., the lattice site). We mainly focus on the special case of equal birth and death rates for reasons of simplicity. For some choices of the model parameters,  the model admits phase-coexistence.  We use mean-field theory (MFT) to analyse the system, supplemented by extensive Monte-Carlo simulation (MCS) studies. Unexpectedly, the MFT predictions on the stationary densities have a rather poor agreement with the corresponding MCS studies results. We attribute this discrepancy to the correlation effects neglected in MFT. We use our MCS studies to estimate the correlation effects, which are finite in the present model.  Quite interestingly, the mean-field phase diagrams with equal birth and death rates for the present model have strong similarities with that in the TASEP with Langmuir kinetics (TASEP-Lk) model with equal attachment-detachment rates, even though the microscopic origins of the two models are very different. Notwithstanding this unexpected similarity, we show that fluctuations have a much stronger impact on the present model than the TASEP-Lk model, making MFT less accurate here. The rest of this article is organised as follows. In Section~\ref{model}, we construct the model that we use. Section~\ref{steady} is devoted to studying the steady states. Next, in Section~\ref{mft}, we set the MFT for this model. Then in Section~\ref{stationary}, we solve the MFT equations to obtain the stationary densities. In Section~\ref{equal-mft}, we solve the MFT equations for equal birth and death rates. Next, in Section~\ref{phase}, we obtain the phase diagram of the model in the plane of the control parameters. In Section~\ref{mft-mcs}, we compare the MFT predictions on the stationary densities with the corresponding MCS results. Then in Section~\ref{mft-inaccuracy}, we analyse our conjecture that correlation effects are responsible for the disagreement between MFT and MCS results, and prodice numerical support to it.  We conclude and summarise in Section~\ref{summ}. %Some additional details are available in Appendix.

\section{Construction of the model}\label{model}

\begin{figure}[htb]
 \includegraphics[width=1.0\columnwidth]{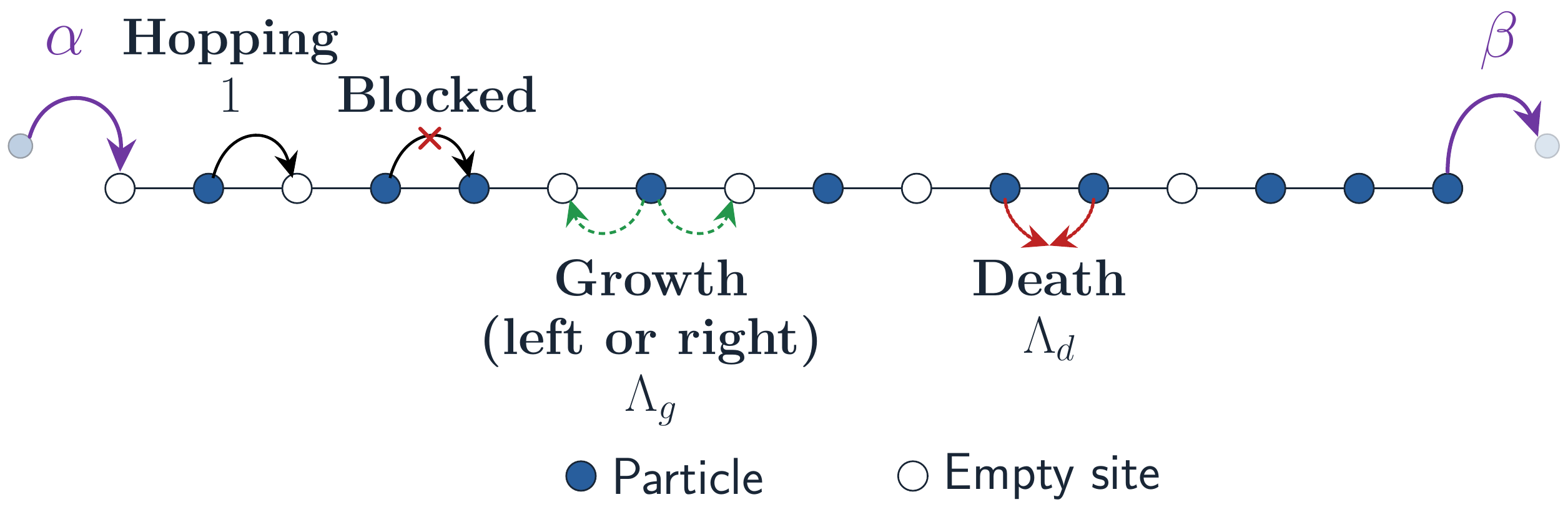}
 \caption{Schematic model diagram. Various microscopic dynamical processes are shown.}\label{fig:Model}
\end{figure}

In this work, we consider a 1D lattice of $L$ sites, which are labelled by $i$, $i=1,..,L$; see Fig.~\ref{fig:Model}.

A site $i$ can have occupation  $n_i=1$ (filled) or $n_i=0$ (empty). The dynamical update rules of this model consist of the following steps:

(a) Hopping: Particles may enter the TASEP lane at the left end ($i=1$) at rate $\alpha\leq 1$, stochastically hop from $i=1$ to $L$ unidirectionally subject to exclusion at rate unity, and leave
the lane at the right end ($i=L$) at rate $\beta\leq 1$~\cite{sm-tasep-atri1}. 

(b) Birth: If a site $i$ is occupied and either site $i-1$ or $i+1$ is vacant, then a particle is created at site $i-1$ or $i+1$ at rate $\Lambda_g>0$.

(c) Death: If sites $i$ and $i-1$ or $i$ and $i+1$ are occupied, then they annihilate each other giving two empty sites at rate $\Lambda_d>0$. This is death due to over population.

The above dynamical rules (a)-(d) define our model.

\section{Steady states }\label{steady}

In this Section, we first construct the MFT for the model defined above, and use that to calculate the stationary densities.

\subsection{Construction of the MFT}\label{mft}

We now construct the MFT equation of motion following the principles of MFT in a pure, open TASEP~\cite{blythe}. MFT entails ignoring spatial correlations. The density $\rho_i$ at site $i$ follows the MFT equation

\begin{eqnarray}
 \frac{\partial \rho_i}{\partial t} &=& \rho_{i-1}(1-\rho_i) - \rho_i(1-\rho_{i+1}) + \frac{\Lambda_g}{2} \bigg[\rho_{i+1}(1-\rho_i) \nonumber \\&&+ \rho_i (1-\rho_{i+i})\bigg] - \frac{\Lambda_d}{2} \bigg[\rho_i \rho_{i-1}+\rho_i\rho_{i+1}\bigg].\label{basic-mft}
\end{eqnarray}
In (\ref{basic-mft}), obviously the birth and death terms do not conserve numbers. Furthermore, the birth and death terms violate the particle-hole symmetry of an open TASEP~\cite{sm-tasep-atri1}. In the thermodynamic limit, we use a quasi-continuous spatial coordinate $x=i/L$ with $L\rightarrow \infty$. Next, to ensure competition between the hopping dynamics of TASEP and the reaction processes (birth and death), we set % $\lambda_g$ and $\lambda_d$ by $1/L$:
\begin{equation}
 \Lambda_g = \frac{\lambda_g}{L},\;\;\Lambda_d = \frac{\lambda_d}{L},
\end{equation}
where $\lambda_d,\lambda_g\sim {\cal O}(1)$.
This is similar to the rescaling of the attachment-detachment rates in TASEP with Langmuir kinetics (Lk) problem; see Refs.~\cite{ef-lktasep-prl,ef-lktasep-pre,tirtha-lk1,tirtha-lk2,sm-tasep-atri2}.
With this and in the steady state setting $\partial\rho/\partial t=0$, we get
\begin{equation}
 \frac{\partial}{\partial x}[\rho(1-\rho)]=\lambda_g\rho - \tilde\lambda \rho^2, \label{mft-steady}
\end{equation}
where $\tilde\lambda \equiv \lambda_g + \lambda_d$. Equation~(\ref{mft-steady}) is the steady state MFT equation, which can be solved subject to suitable boundary conditions. It allows us to extract an MFT expression for the steady state current $J$ given by
\begin{equation}
J = \rho(1-\rho),
\end{equation}
where any boundary layers are not considered in the thermodynamic limit.

\subsection{Stationary densities}\label{stationary}

We now solve (\ref{mft-steady}) to calculate the stationary densities, subject to the usual boundary conditions $\rho(0)=\alpha$ and $\rho(1)=1-\beta$. We note from (\ref{mft-steady}) that $\rho(x)=m$, a constant, is a solution to (\ref{mft-steady}), provided
\begin{equation}
 \lambda_g m - \tilde \lambda m^2 = 0,
\end{equation}
giving
\begin{equation}
 m= \frac{\lambda_g}{\tilde\lambda},\label{m-sol}
\end{equation}
assuming $m\neq 0$. This solution $\rho(x)=m$ can be more or less or equal to 1/2 and clear {\em does not} satisfy any of the boundary conditions at $x=0,1$. The latter property is reminiscent of the MC phase in an open TASEP, and hence $\rho(x)=m\equiv \rho_\text{GMC}(x)$ is a generalised MC phase; see also Refs.~\cite{ef-lktasep-prl,ef-lktasep-pre,tirtha-lk1,tirtha-lk2} in the context of the TASEP-Lk model. However, this is {\em not} the only solution to $\rho(x)$. The general solution for $\rho(x)$ may be written as
\begin{equation}
 \rho(x)=\phi(x)+m,\label{gen-sol}
\end{equation}
where $m$ is as given by (\ref{m-sol}). Here, $\phi(x)$ is yet to be determined, $x$-dependent function, giving an $x$-dependent $\rho(x)$. By substituting (\ref{gen-sol}), we get the equation for $\phi(x)$:
\begin{equation}
 \frac{\partial}{\partial x} (A\phi -\phi^2)= - \phi (\lambda_g + \tilde \lambda\phi),\label{phi-eq}
\end{equation}
where $A\equiv 1-2\lambda_g/\tilde\lambda$. We now solve (\ref{phi-eq}), which can be recast as
\begin{equation}
 \frac{Ad\phi-2\phi d\phi}{\phi(\lambda_g + \tilde\lambda\phi)} = -dx.
\end{equation}
Integrating, we obtain
\begin{equation}
 \ln \phi^{A/\lambda_g}-\ln \left(\phi + \frac{\lambda_g}{\tilde\lambda}\right)^{\frac{A}{\lambda_g}+\frac{2}{\tilde\lambda}} = -x+C,
 \end{equation}
where $C$ is a constant of integration. Simplifying,
\begin{equation}
 \phi^{\frac{A}{\lambda_g}}{\left(\phi + \frac{\lambda_g}{\tilde\lambda}\right)^{\frac{A}{\lambda_g}+\frac{2}{\tilde\lambda}}}=Be^{-x},\label{phi-gen-sol}
\end{equation}
where $B$ is also a constant of integration. Equation~(\ref{phi-gen-sol}) gives an implicit solution of $\phi(x)$.  Constant $B$ can be evaluated from the boundary conditions on $\rho(x)$ at $x=0,1$. There are two possible cases:

(i) Injection limited solution: This is the analogue of the LD phase in a pure, open TASEP~\cite{blythe,kolo}. Use the boundary condition $\rho(x=0)=\alpha$. This gives after using (\ref{gen-sol})
\begin{equation}
 B_\text{in}= \frac{\phi_0^{A/\lambda_g}}{\left(\phi_0 + \frac{\lambda_g}{\tilde\lambda}\right)^{A/\lambda_g + 2/\tilde\lambda}},
\end{equation}
where 
\begin{equation}
\phi_0\equiv \phi(x=0)=\rho (x=0) - m = \alpha -m.
\end{equation}

(ii) Extraction limited solution:  This is the analogue of the HD phase in a pure, open TASEP~\cite{blythe,kolo}. Use the boundary condition $\rho(x=1)=1-\beta$. Once again, using (\ref{gen-sol})
\begin{equation}
 B_\text{ex}=\frac{e\phi_1^{A/\lambda_g}}{\left(\phi_1 + \frac{\lambda_g}{\tilde\lambda}\right)^{A/\lambda_g + 2/\tilde\lambda}},
\end{equation}
where 
\begin{equation}
 \phi_1 \equiv \phi(x=1)= \rho(x=1)-m.
\end{equation}

Therefore, we find that there are three basic stationary solutions for $\rho(x)$, which are $\rho_\text{in}(x),\rho_\text{ex}(x)$ and $\rho_\text{GMC}$, the last one being a constant.

For reasons of convenience in analytical manipulations and explicit calculations, we consider the special case with $\lambda_g=\lambda_d=\lambda$, giving $\tilde\lambda=2\lambda$. This allows us to obtain explicit analytical solution for $\phi(x)$ from the MFT. We then compare that with MCS studies results. These are presented below.

\subsection{$\lambda_g=\lambda_d=\lambda$} \label{equal-mft}

In this case, $m=1/2$ and $A=1-2\lambda_g/\tilde \lambda=0$. From (\ref{phi-gen-sol}), we then obtain
\begin{equation}
 \left(\phi + \frac{1}{2}\right)^{2/\tilde\lambda} = B^{-1}e^x,
\end{equation}
giving
\begin{equation}
 \phi(x) = \overline B \exp(\tilde\lambda x/2) - \frac{1}{2}, \label{phi-spl-soln}
\end{equation}
giving an explicit solution of $\phi$ as a function of $x$. Here, $\overline B\equiv B^{(-\tilde\lambda/2)}$ is a constant.
We then get
\begin{equation}
 \rho(x)= \phi(x)+m=\overline B \exp(\tilde\lambda x/2),
\end{equation}
in addition to $\rho(x)=m=1/2$ (i.e., $\phi(x)=0$) that is identical to the MC phase solution of an isolated open TASEP.

We now use the boundary conditions on $\rho(x)$ to calculate $\overline B$. At $x=0$, $\rho(x=0)=\alpha$, giving $\overline B=\alpha$. Thus
\begin{equation}
\rho_\text{in}=\alpha e^{\tilde\lambda x/2}. \label{rho-in}
\end{equation}
This is a {\em growing} solution in $x$. This in turn means for this solution $\rho_\text{in}(x>0)>\alpha$ in general. 
If $\alpha<1/2$, $\rho_\text{in}(x)\leq 1/2$ in the range $0\leq x\leq -(2/\tilde\lambda)\ln (2\alpha)$, reminiscent of the LD phase in pure open TASEP~\cite{blythe,kolo}. Given that $\rho_\text{in}(x)$ is controlled by the ``left'' boundary condition $\rho(x=0)=\alpha$, we are thus tempted to identify $\rho_\text{in}(x)=\rho_\text{LD}(x)$, a generalised LD phase stationary density profile. There are, however, two major differences vis-\`a-vis the usual LD phase density in an open TASEP, {\em viz.} (i) it is space dependent here, whereas in a pure open TASEP, it is constant in the bulk; (ii) as a consequence of the space-dependence of $\rho_\text{LD}(x)$, the LD phase may span only a part of the TASEP lane, depending upon $\alpha$; see also below.

Next, at $x=1$, $\rho(x=1)=1-\beta$, giving $\overline B=(1-\beta)\exp(-\lambda)$. Thus
\begin{equation}
 \rho_\text{ex}=(1-\beta)e^{\lambda (x-1)}.\label{rho-ex}
\end{equation}
This is also a growing solution in $x$. This in turn means that for this solution $\rho_\text{ex}(x<1)<1-\beta$ in general. In fact, if $\beta<1/2$, i.e., $1-\beta>1/2$, $\rho_\text{ex}(x)\leq 1/2$ in the range $1-(2/\tilde\lambda)\ln [2(1-\beta)]\leq x\leq 1$, which reminds us of the HD phase in pure open TASEP. Given that $\rho_\text{ex}(x)$ is controlled by the ``right'' boundary condition $\rho(x=1)=\beta$, we are thus tempted to identify $\rho_\text{ex}(x)=\rho_\text{HD}(x)$, a generalised HD phase stationary density profile. Similar to the LD phase, there are, however, two major differences vis-\`a-vis the usual HD phase density in an open TASEP, {\em viz.} (i) it is space dependent here, whereas in a pure open TASEP, it is constant in the bulk; (ii) as a consequence of the space-dependence of $\rho_\text{HD}(x)$, the HD phase may span only a part of the TASEP lane, depending upon $\beta$; see also below. Notice the similarity between the three basic solutions here and the three basic solutions of the TASEP-Lk problem with equal attachment-detachment rate~\cite{ef-lktasep-prl,ef-lktasep-pre}. Similar to the TASEP-Lk problem, the space-dependent nature of $\rho_\text{LD}(x)$ and $\rho_\text{HD}(x)$, this model allow for mixed phases, leading to phase coexistence.  However, the $x$-dependence of $\rho_\text{LD}(x)$ and $\rho_\text{HD}(x)$ is nonlinear in $x$ in the present model, unlike in the TASEP-Lk model, where such dependence is linear with equal attachment-detachment limit.

%Phenomenological approach to improve MFT

\subsection{Phase diagram}\label{phase}

With the three basic solutions $\rho_\text{LD}(x), \rho_\text{HD}(x), \rho_\text{MC}=1/2$, we can now construct the MFT phase diagram in the $\alpha-\beta$ plane and mark the various phases. The phase boundaries are determined by equality of the currents in the different phases. This holds because, although the birth-death processes make the model nonconserved, the probability of occurrence of birth and death at a given site is vanishingly small in the limit of $L\rightarrow\infty$. This ensures that the steady state currents are {\em locally} conserved; see also Refs.~\cite{ef-lktasep-pre}. As mentioned above, $\rho_\text{LD}(x)$, being a rising function, may exceed 1/2 at $x<1$, which should mark the beginning of the MC phase. Equating the currents in the LD phase and MC phase at $x=x_\alpha$, we get
\begin{equation}
 x_\alpha= -\frac{1}{\tilde\lambda}\ln (2\alpha),\label{ld-mc}
\end{equation}
such that the system is in the LD phase for $0\leq x\leq x_\alpha$ and is in the MC phase between $x_\alpha<x<1$, giving the LD-MC phase. Clearly,  $x_\alpha=1$ marks the changeover from the LD to LC-MC phase. Setting $x_\alpha=1$ in (\ref{ld-mc}) above, we get
\begin{equation}
 \alpha = \frac{1}{2}e^{-\tilde\lambda} \label{ld-mc-boun}
\end{equation}
as the phase boundary between the LD and LD-MC phase in the $\alpha-\beta$ plane. On the other hand, $x_\alpha=0$ sets the condition for the changeover from LD-MC to MC phase. Setting $x_\alpha=0$ in (\ref{ld-mc}), we get
\begin{equation}
 \alpha=\frac{1}{2},
\end{equation}
which is identical to the LD-MC phase boundary in the $\alpha-\beta$ plane in a pure, open TASEP. Here, it is however the phase boundary between a distinct set of phases, {\em viz.}, LD, pure phase, and LD-MC, a mixed phase.

Next, we consider the fact that the solution $\rho_\text{HD}(x)$ can be less than 1/2 at some $x$. To proceed quantitatively, let us assume at $x=x_\beta$, $\rho_\text{HD}(x_\beta)=1/2$. This gives
\begin{equation}
 (1-\beta)e^{\tilde\lambda(x_\beta-1)}=\frac{1}{2},
\end{equation}
which gives
\begin{equation}
 x_\beta=1-\frac{1}{\lambda}\ln [2(1-\beta)].
\end{equation}
The boundary between the MC and HD-MC phase is determined by $x_\beta=1$, giving
\begin{equation}
 \ln [2(1-\beta)]=\tilde\lambda,
\end{equation}
which implies
\begin{equation}
 \beta=1=\frac{1}{2}e^{\tilde\lambda}.
\end{equation}
 Furthermore, mixed phase HD-MC to pure phase HD transition is given by the condition $x_\beta=0$, which gives
\begin{equation}
 \ln [2(1-\beta)]= 0,
\end{equation}
giving
\begin{equation}
 \beta=\frac{1}{2}.
\end{equation}
Although this condition is identical to the one at the HD-MC transition in a pure, open TASEP, here is refers to a different transition, between a pure HD phase and a mixed HD-MC phase.

We have thus found that the model admits several mixed phases, LD-MC and HD-MC. Similar to the TASEP-Lk problem, it is possible to have three-phase coexistence if
\begin{equation}
 x_\alpha<x_\beta
\end{equation}
or
\begin{equation}
 -\frac{1}{\tilde\lambda}\ln (2\alpha)< 1- \frac{1}{\tilde\lambda}\ln [2(1-\beta)].\label{3-phase}
\end{equation}
If the condition (\ref{3-phase}) is satisfied, between $\rho (x)<1/2$ for $0\leq x< x_\alpha$, $\rho)x)=1/2$ in the range $x_\alpha \leq x\leq x_\beta$ and $\rho(x)>1/2$ for $x_\beta< x\leq 1$, giving a three-phase coexistence. %This need not hold for any $\alpha,\beta,\lambda$.

Three phase coexistence naturally requires $x_\alpha<x_\beta$. For $x_\alpha>x_\beta$, one instead has a two-phase coexistence - LD-HD phase, which is a localised domain wall (LDW). Since the nonconserving reaction terms are locally vanishingly small in the limit $L\rightarrow \infty$, the TASEP hopping current should be locally conserved. This is similar to Ref.~\cite{ef-lktasep-pre,sm-tasep-atri2}. This gives
\begin{equation}
 \rho_\text{LD}(x_w)+\rho_\text{HD}(x_w)=1,
\end{equation}
where $x_w$ is the position of the LDW. We then find
\begin{equation}
 x_w = \frac{1}{\tilde \lambda}\ln \bigg|\frac{1}{\alpha + (1-\beta)e^{-\tilde\lambda}}\bigg|.\label{x-w}
\end{equation}

\begin{figure}[htb]
 \includegraphics[width=0.5\columnwidth]{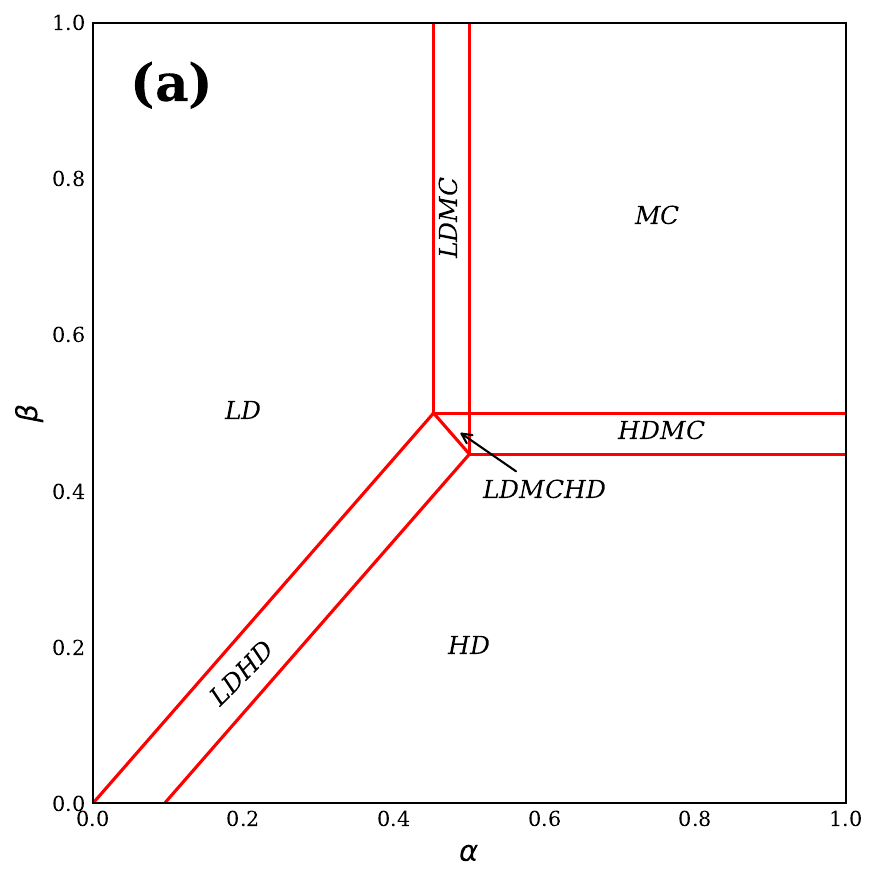}\hfill \includegraphics[width=0.5\columnwidth]{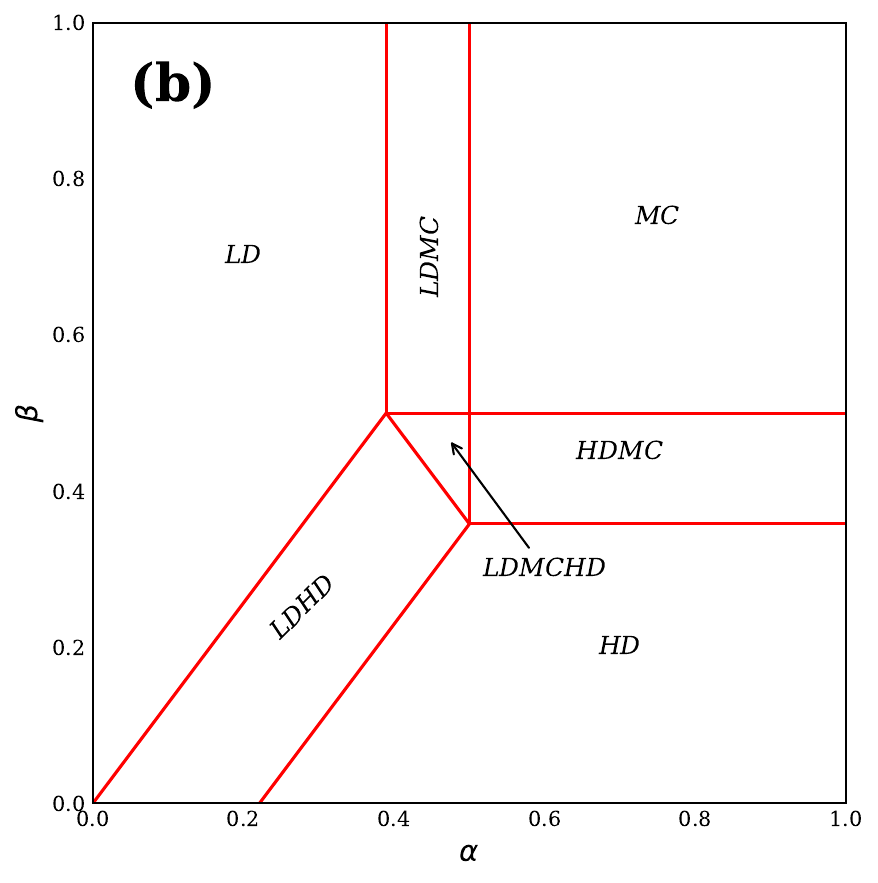}\\
 \includegraphics[width=0.5\columnwidth]{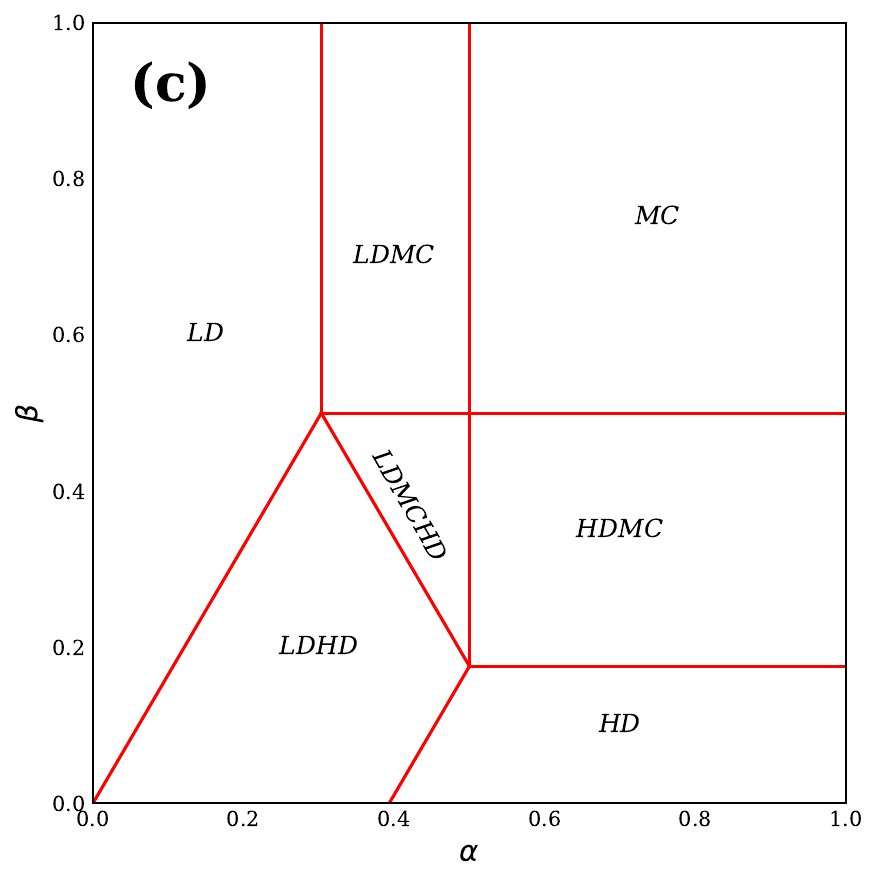}\hfill \includegraphics[width=0.5\columnwidth]{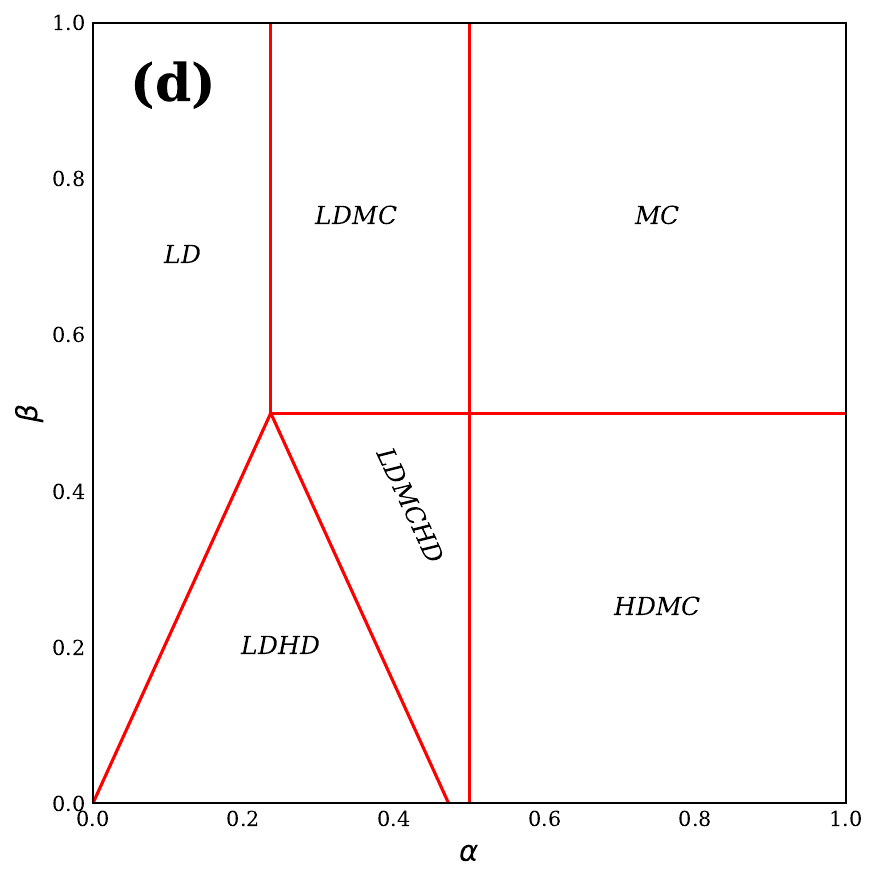}\\
 \includegraphics[width=0.5\columnwidth]{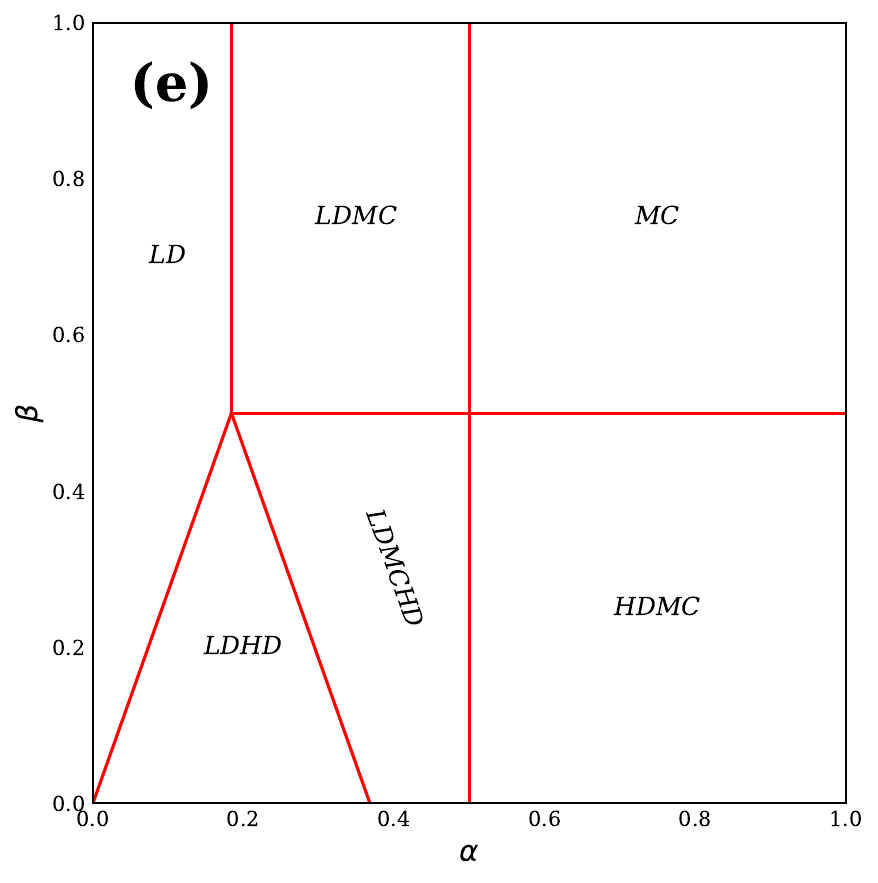}
 \caption{MFT phase diagrams in the $\alpha-\beta$ plane in (a)-(e) for $\lambda=0.1, 0.25, 0.5, 0.75,1.0$ respectively.}\label{phase-diag}
\end{figure}

See Fig.~\ref{phase-diag} for the MFT phase diagrams in the $\alpha-\beta$ plane.

Notice that with increasing $\lambda$ the mixed phases occupy larger areas in the phase diagram. These mixed phases do not exist in the pure TASEP limit with $\lambda=0$. Thease features can be explanied from the MFT equation~\eqref{mft-steady}, which in the limit $\lambda_g=\lambda_d=\lambda$ becomes
\begin{equation}
 \frac{\partial}{\partial x}[\rho(1-\rho)]= \lambda\rho (1-2\rho).\label{mft-equal-lambda}
\end{equation}
The rhs of (\ref{mft-equal-lambda}), the ``reaction source'', is positive for $\rho<1/2$, negatuve for $\rho>1/2$ and vanishes when $\rho=1/2$. Therefore, at the level of this simple MFT, the reaction processes tend to drive the density towards 1/2: In the LD phase, the reactions tend to increase the density towards 1/2, whereas in the HD phase, reactions tend to drive down the density towards 1/2. This suggests that the uniform LD or HD phase in a pure TASEP  can get modified to produce the mixed phases. Since the reaction term in (\ref{mft-equal-lambda}) scales with $\lambda$, the extent of the mixed phases increase with $\lambda$. Since the MC phase density $\rho=1/2$ is unaffected by the reaction term in (\ref{mft-equal-lambda}), the MC phase in the MFT phase diagram should remain unaffected by the reaction. This is consistent with the phase diagrams in Fig.~\ref{phase-diag}.

Just as the basic MFT soluations for the stationary solutions in the present problem have a degree of correspondence with those in the TASEP-Lk problem, the phase diagrams in two problems are also similar. Nonetheless, there are significant sifference at the level of the microscopic dynamics. In the TASEP-Lk problem, the attachment-detachment processes are purely local, being unaffected by the occupancy of the neighbouring sites. In the present model, reactions processes involve neighbouring sites, which develop correlations, which in turn ultimately make the MFT predictions not agreeing with the MCS results; see below.

\section{Stationary densities - MFT versus MCS}\label{mft-mcs}

\begin{figure}[htb]
\includegraphics[width=0.9\columnwidth]{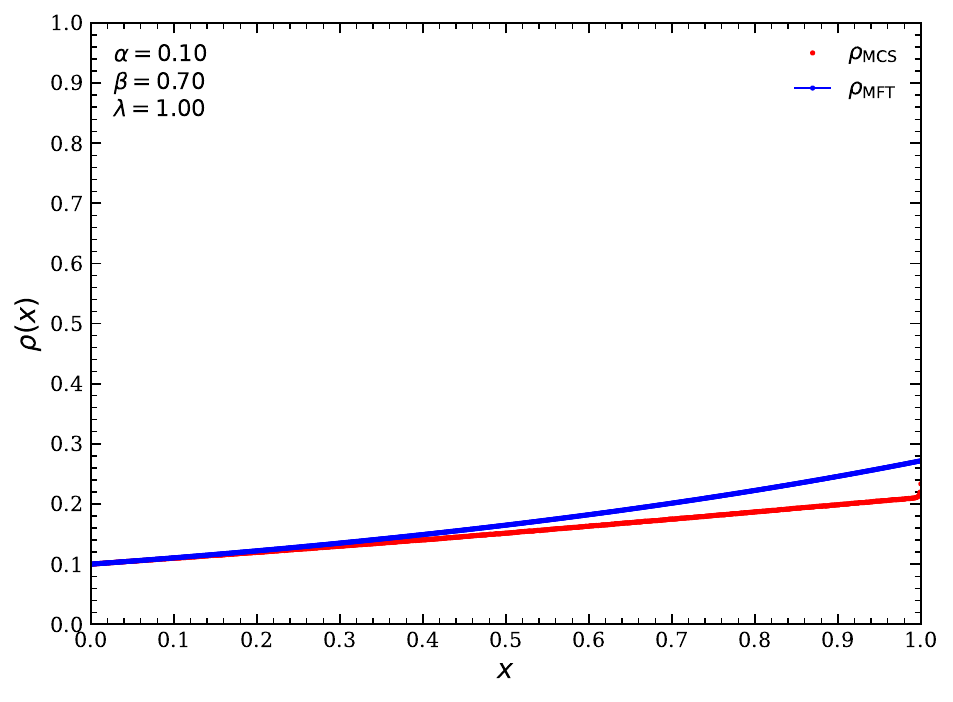}\\
\includegraphics[width=0.9\columnwidth]{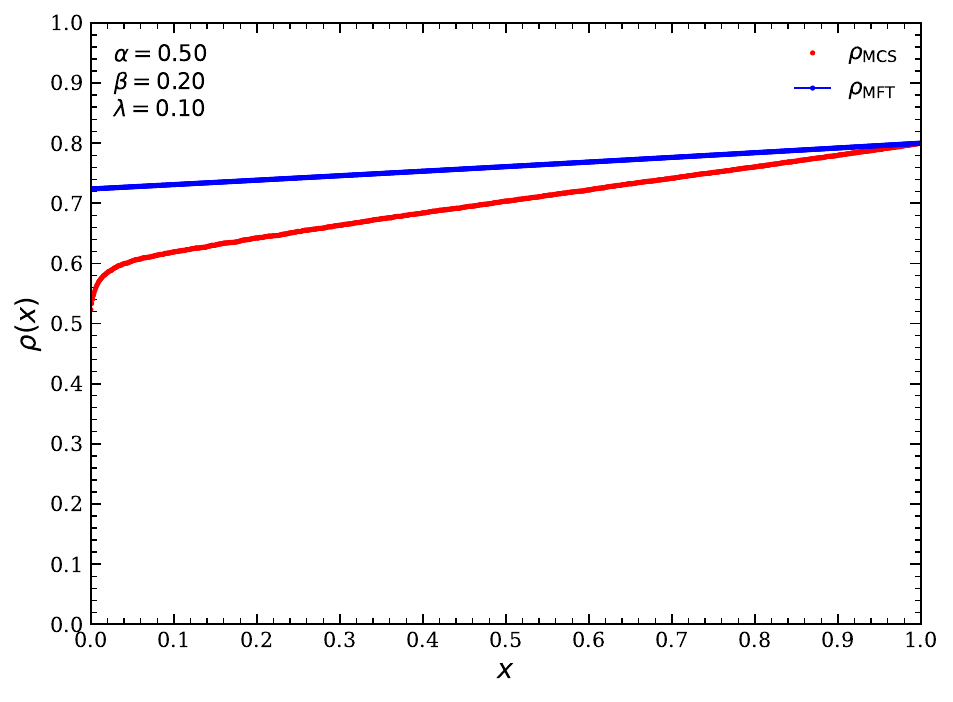}
\caption{Stationary densities in the LD and HD phases obtained from MFT (blue line) and MCS (red points). MFT predictions for both the LD (top) and HD (bottom) phase stationary densities are consistently above the MCS results. See text.}\label{ld-hd}
\end{figure}

In this Section, we show the MFT stationary density profiles in the different phases, and compare them with the corresponding MCS results; see the representative plots for the stationary densities in the LD and HD phases in Fig.~\ref{ld-hd}. We connect the discrepancy between the MFT and MCS results to the correlation effects.

%\begin{widetext}

We find that there is a significant disagreement between the MFT predictions on the stationary densities and the corresponding MCS results; see Fig.~\ref{ld-hd}. This is rather unexpected, because in an open TASEP-like model there is no global number conservation. As a result, the density fluctuations are short-ranged, which in turn should make MFT reasonably accurate. This expectation usually holds good in different open TASEP or its variants, including TASEP-Lk~\cite{ef-lktasep-prl}. Significant deviations are usually found only in some models with a strict number conservation, e.g., \cite{tirtha-lk1,sourav1,sourav2}. The unexpected disagreements between MFT and MCS results in the present model require careful considerations, which we discuss below.

We now argue that  the correlation effects, neglected in the derivation of the MFT Eq.~\eqref{basic-mft}, are responsible for this disagreement between the MFT and MCS results.
To investigate the correlation effects, we first study the system in a limit in which the dynamics is overwhelmingly dominated by the reaction part. This is achieved by {\em removing} the $1/L$-scaling that was introduced to the birth and death rates to ensure competition between  the reaction process and the TASEP hopping dynamics. If our argument explaining the lack of accuracy of MFT is indeed correct, then that should affect even in this extreme case of the reaction-dominated dynamics. In this limit, neglecting the TASEP hopping dynamics, which is now irrelevant in the bulk, the MFT predicts that the bulk density should be
\begin{equation}
 \rho_\text{bulk}^\text{MFT}= \frac{\lambda_g}{\lambda_d+\lambda_g}\label{test-bulk}.
\end{equation}

\begin{figure}[htb]
 \includegraphics[width=0.9\columnwidth]{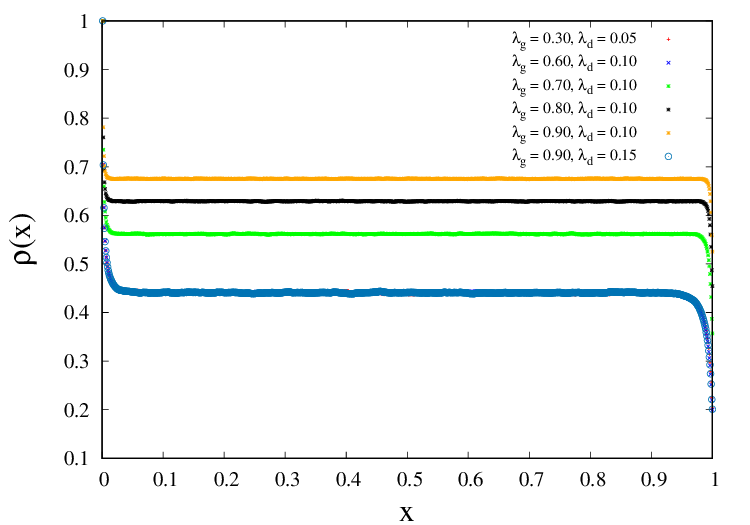}
 \caption{Stationary densities $\rho_\text{bulk}$ in the limit in which the reactions dominate over the TASEP hopping for large systems for various $\lambda_g,\lambda_d$ in the MCS studies. We find that $\rho_\text{bulk}<\rho_\text{bulk}^\text{MFT}$. See text.}\label{mcs-test-bulk}
\end{figure}

To verify the result~\eqref{test-bulk}, we study the model with MCS, after removing the $1/L$-scaling, as discussed above. The MCS results are shown in Fig.~\ref{mcs-test-bulk}.
We made the following observations from Fig.~\ref{mcs-test-bulk}. First of all, the bulk density $\rho_\text{bulk}$ remains unchanged if both $\lambda_g$ and $\lambda_d$ are enhanced or decreased by the same factor. For instance in Fig.~\ref{mcs-test-bulk}, the bulk density is same for $(\lambda_g = 0.30, \lambda_d = 0.05)$, $(\lambda_g =0.60, \lambda_d = 0.10)$ and $(\lambda_g = 0.90, \lambda_d = 0.15)$, implying that if $\lambda_g,\lambda_d$ change keeping the ratio $\lambda_g/(\lambda_d+\lambda_g)$ unchanged. However, the MCS density in the bulk $\rho_\text{bulk}< \rho_\text{bulk}^\text{MFT}= \lambda_g/(\lambda_d+\lambda_g)$, violating the MFT prediction. Thus the MFT predictions appear to overestimate the stationary densities, as we also find in Fig.~\ref{ld-hd}.

%In fact, we find that a nonzero $\rho_\text{bulk}$ is obtained provided $\lambda_g/(\lambda_g+\lambda_d) \geq 6/7$; otherwise, if $\lambda_g/(\lambda_g+\lambda_d) < 6/7$, $\rho_\text{bulk}$ is vanishingly small.

%This shows that $\lambda_g^\text{eff}$, the effective $\lambda_g$, which controls $\rho_\text{bulk}$ turns negative if $\lambda_g/(\lambda_g+\lambda_d) < 6/7$. This suggest that $\lambda_g^\text{eff}<\lambda_g$. %How much less it is than $\lambda_g$ can be estimated as follows. Defining $\lambda_g^\text{eff}$ by
%\begin{equation}
% \rho_\text{bulk}\equiv \frac{\lambda_g^\text{eff}}{\lambda_g + \lambda_d}.\label{lambda-eff-def}
%\end{equation}
%From our MCS studies in Fig.~\ref{mcs-test-bulk}, we phenomenologically write
%\begin{equation}
% \rho_\text{bulk}=\frac{ \lambda_g}{\lambda_d+\lambda_g}-a,\label{pheno}
%\end{equation}
%where $a$ is a free parameter.
%Combining \eqref{lambda-eff-def} with \eqref{pheno}, we write a phenomenological relation
%\begin{equation}
% \lambda_g^\text{eff}= \lambda_g(1-a)-a\lambda_d. \label{lambda-eff-soln}
%\end{equation}

%[Plot of comparison of the above formula with the TASEP-less case bulk densities]

\section{Why is MFT inaccurate here?}\label{mft-inaccuracy}

We now discuss why MFT is quantitatively inaccurate, in particular why it {\em overestimates} the staionary densities. The derivation of the MFT Eq.~\eqref{basic-mft} involves replacing the death term
$\langle n_i n_{i\pm1}\rangle$ by $\rho^2$, where $\rho=\langle n_i\rangle$. This factorisation becomes qestionable in 1D for the following reason. The birth process tends to create clusters of occupied sites along the TASEP lane. Once a cluster is formed, the birth process can take place only along the edges of the cluster. In contrast, the death process can take place inside the bulk of the cluster. This suppresses the birth events, effectively suppressing the birth rate, relative to the death rate. As a result, the MCS gives a lower steady state density in comparison with the MFT prediction. Equivalently, it is expected that $\langle n_i n_{i\pm 1}\rangle >\rho^2$, giving the true loss term $\lambda_d \langle n_i n_{i\pm 1}\rangle> \lambda_d\rho^2$, its mean field value. This is another way of seeing that the ``true'' steady state density should be lower than that predicted by MFT. We now make quantitative estimates of $P=\langle n_i n_{i+1}\rangle$ relative to $Q=\rho^2$ from our MCS results in the stationary states in pure TASEP and in the present model. If MFT works well, we expect $P/Q\approx 1$, else $P/Q\neq 1$.

The actual reaction term on the rhs of (\ref{mft-equal-lambda}) is
\begin{equation}
 R_\text{actual}= \lambda (\rho - 2P),
\end{equation}
whereas the MFT reaction term is
\begin{equation}
 R_\text{MFT}= \lambda (\rho-2\rho^2).
\end{equation}
Clearly, if $P> (<) \rho^2\equiv Q$, the actual death rate is larger (smaller) than one appears in the MFT equation~\eqref{mft-equal-lambda}. If it is larger (smaller), then the actual stationary density should be smaller (larger) than that predicted by MFT. We check this numerically. We calculate $P/Q$ in the different phases and also in pure TASEP (without any reaction). If the factorisation works, then $P/Q\approx 1$ in our MCS studies, else $P/Q\neq 1$.

\begin{figure}[htb]
 \includegraphics[width=\columnwidth]{Rohn-mcs_plots-2.pdf}\\
\includegraphics[width=0.5\textwidth]{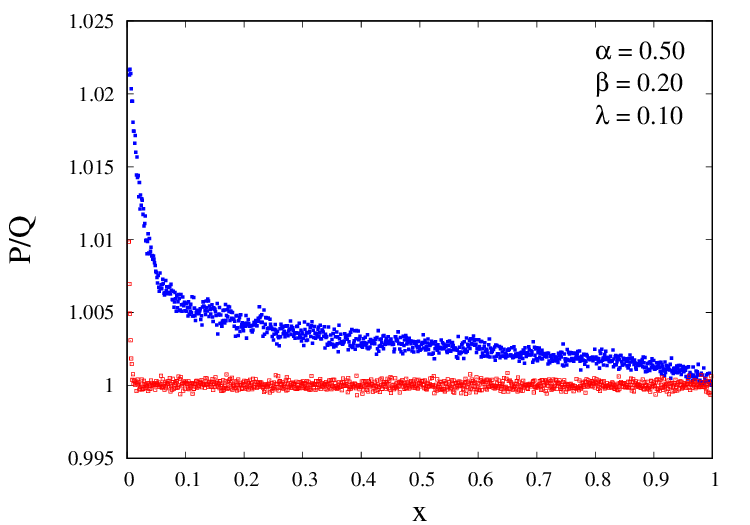}
\caption{(top) Stationary density for $\alpha=0.5,\beta=0.2,\lambda=0.1$ from MFT (blue line) and MCS (red points). MFT density is higher than MCS density. (bottom) Plot of $P/Q$ versus $x$ for pure TASEP (red points) and TASEP with reaction (blue points). For the latter, $P/Q>1$ and decreases with $x$. This suggests an effective death rate larger than $\lambda$ that decreases with $x$, which is consistent with the top figure. }\label{fig-1}
\end{figure}

\begin{figure}[htb]
 \includegraphics[width=\columnwidth]{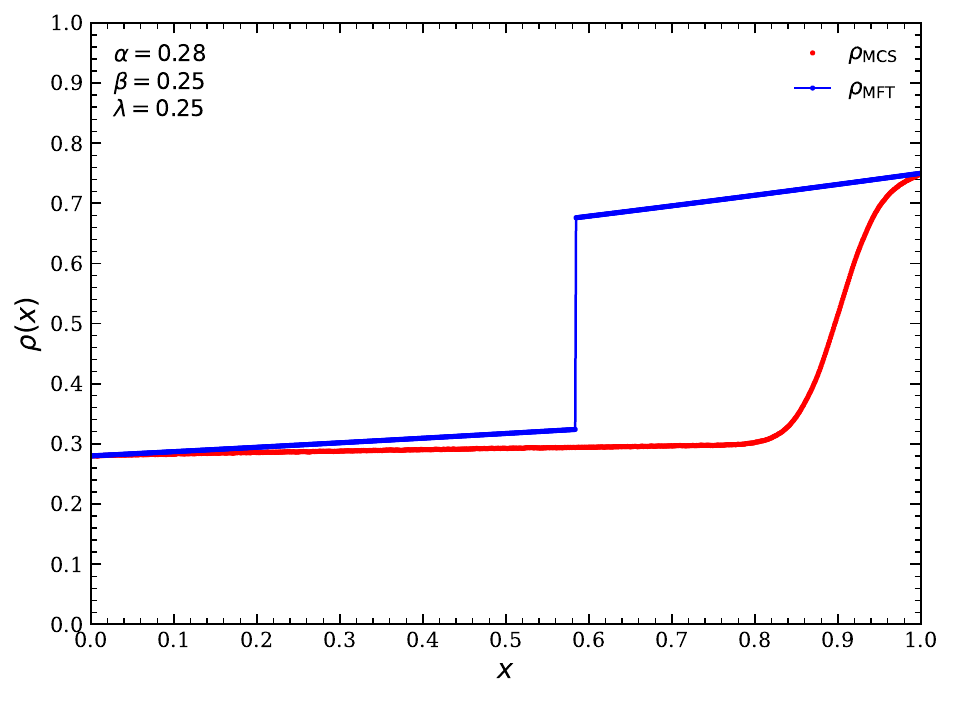}\\
 \includegraphics[width=0.5\textwidth]{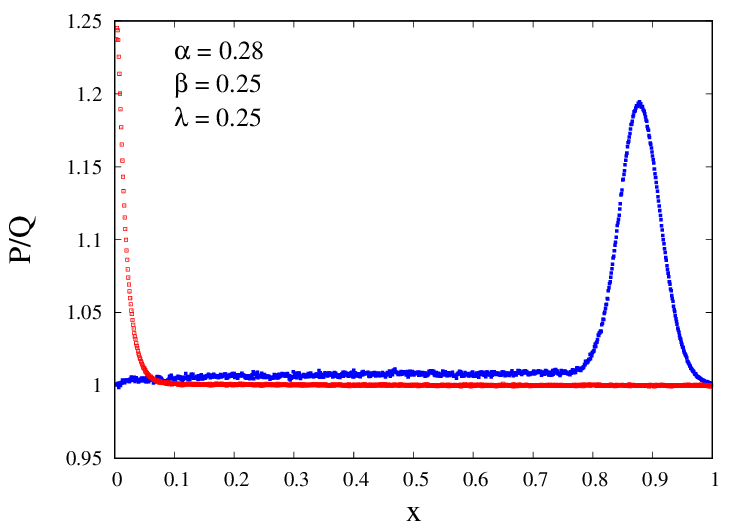}
 \caption{(top) Stationary density for $\alpha=0.28,\beta=0.25,\lambda=0.25$ from MFT (blue line) and MCS (red points). MFT density is higher than MCS density. (bottom) Plot of $P/Q$ versus $x$ for pure TASEP (red points) and TASEP with reaction (blue points). For the latter, $P/Q>1$ and has a peak towards the exit end  ($x=1$). This suggests an effective death rate larger than $\lambda$ that has a peak towards the exit end  ($x=1$), which is consistent with the top figure.}\label{fig-2}
 \end{figure}

\begin{figure}[htb]
 \includegraphics[width=\columnwidth]{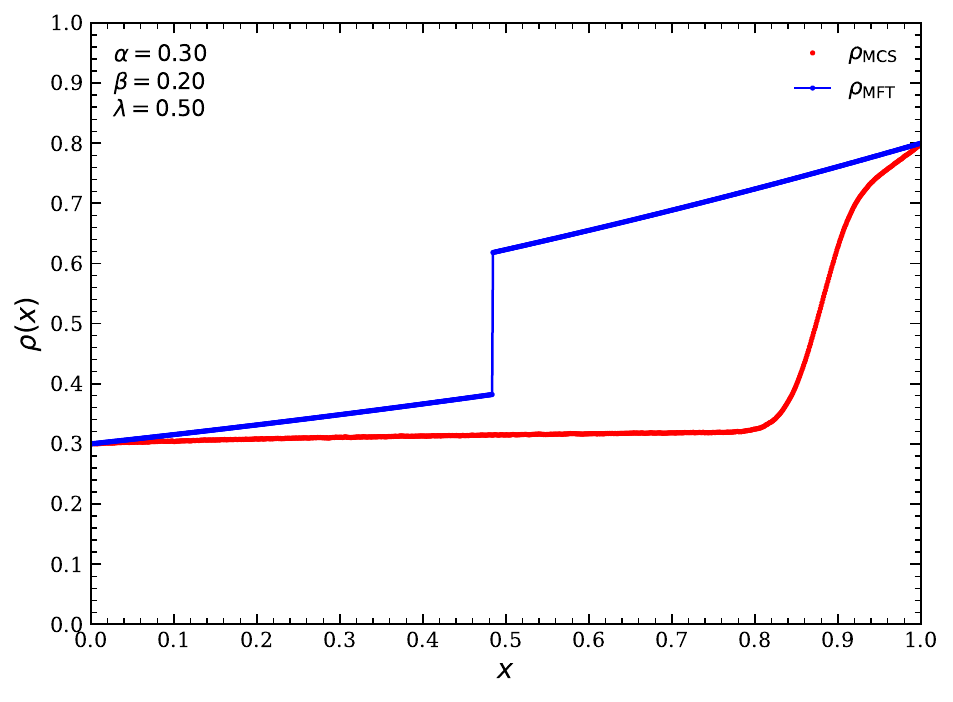}\\
 \includegraphics[width=0.5\textwidth]{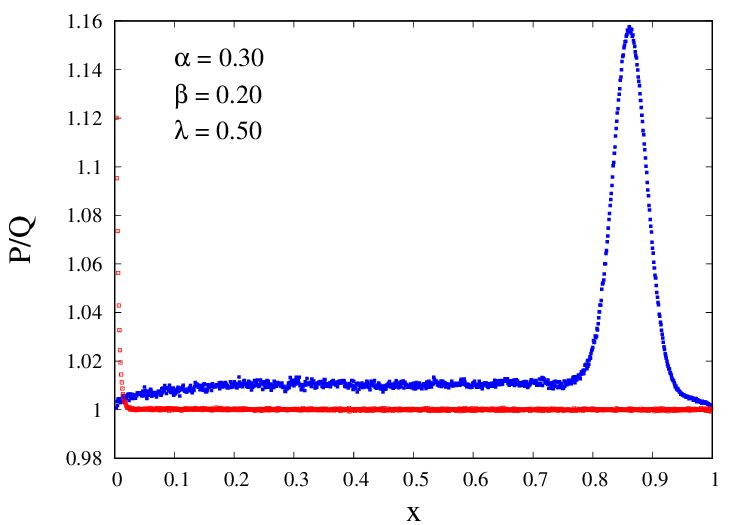}
 \caption{(top) Stationary density for $\alpha=0.3,\beta=0.2,\lambda=0.5$ from MFT (blue line) and MCS (red points). MFT density is higher than MCS density. (bottom) Plot of $P/Q$ versus $x$ for pure TASEP (red points) and TASEP with reaction (blue points). For the latter, $P/Q>1$ and has a peak towards the exit end  ($x=1$). This suggests an effective death rate larger than $\lambda$ that has a peak towards the exit end  ($x=1$), which is consistent with the top figure.}\label{fig-3}
 \end{figure}

\begin{figure}[htb]
\includegraphics[width=\columnwidth]{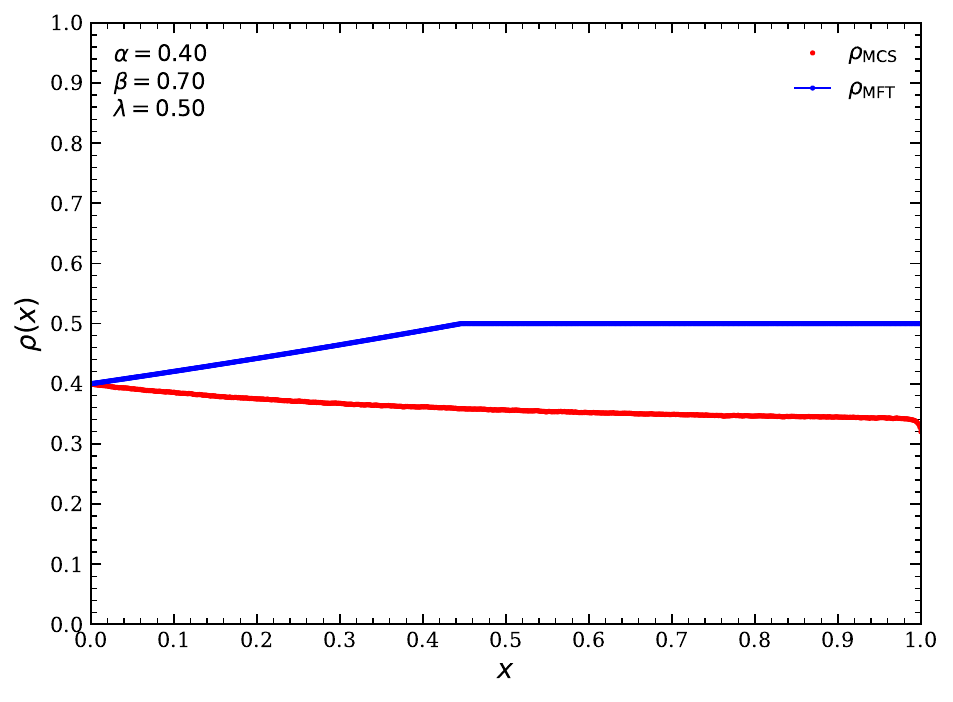}\\
 \includegraphics[width=0.5\textwidth]{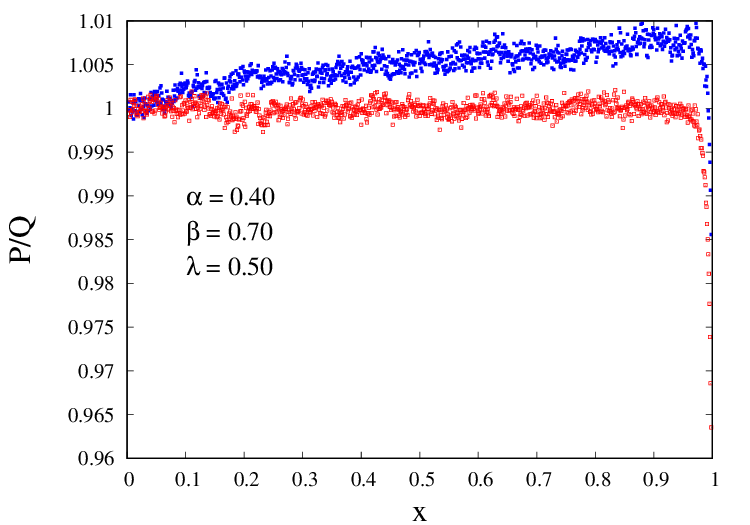}
 \caption{(top) Stationary density for $\alpha=0.4,\beta=0.7,\lambda=0.5$ from MFT (blue line) and MCS (red points). MFT density is higher than MCS density. (bottom) Plot of $P/Q$ versus $x$ for pure TASEP (red points) and TASEP with reaction (blue points). For the latter, $P/Q>1$ and grows with $x$. This suggests an effective death rate larger than $\lambda$ that grows with $x$, which is consistent with the top figure.}\label{fig-4}
 \end{figure}

 \begin{figure}[htb]
 \includegraphics[width=\columnwidth]{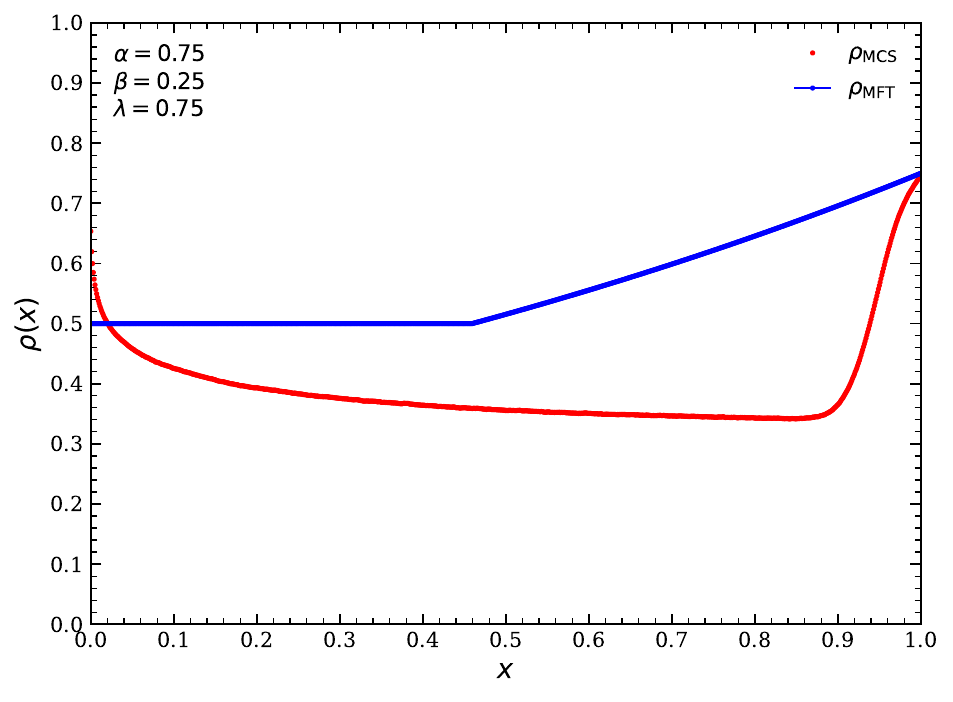}\\
 \includegraphics[width=0.5\textwidth]{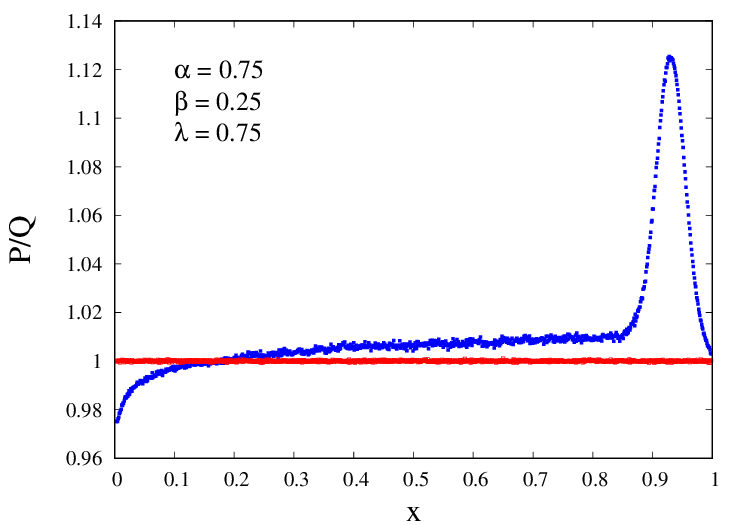}
 \caption{(top) Stationary density for $\alpha=0.75,\beta=0.25,\lambda=0.75$ from MFT (blue line) and MCS (red points). MFT density is higher than MCS density. (bottom) Plot of $P/Q$ versus $x$ for pure TASEP (red points) and TASEP with reaction (blue points). For the latter, $P/Q>1$ and has a large peak for  $x$ near 1. This suggests an effective death rate larger than $\lambda$ that has a large value for $x$ near 1, which is consistent with the top figure.}
 \label{fig-5}
  \end{figure}

 \begin{figure}[htb]
  \includegraphics[width=\columnwidth]{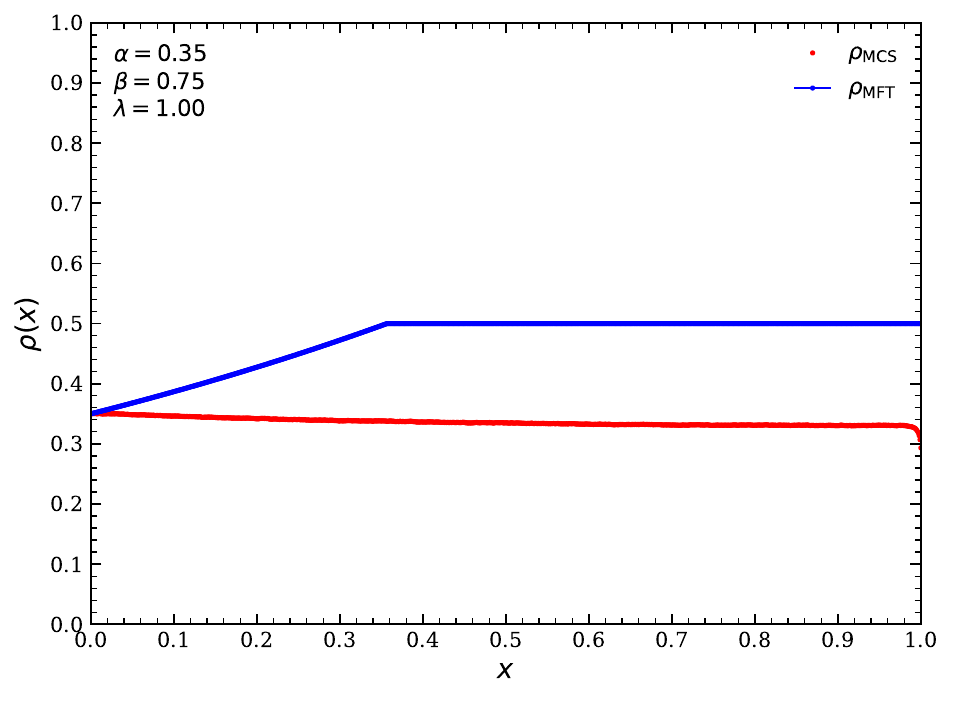}\\
 \includegraphics[width=0.5\textwidth]{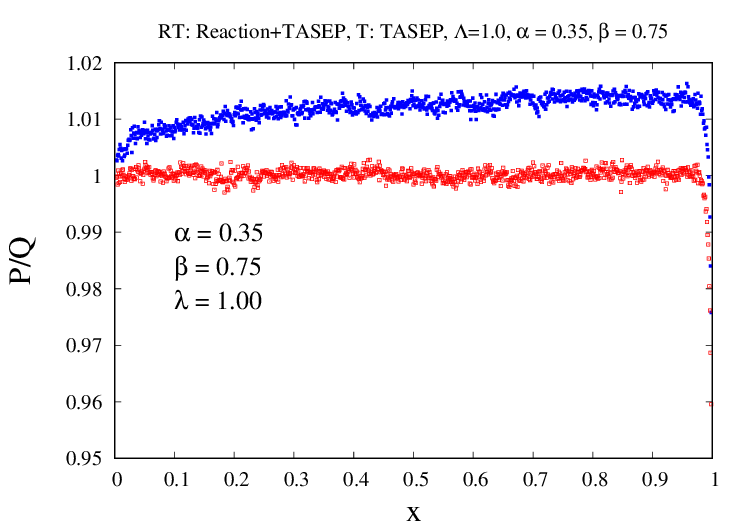}
 \caption{(top) Stationary density for $\alpha=0.35,\beta=0.75,\lambda=1.0$ from MFT (blue line) and MCS (red points). MFT density is higher than MCS density. (bottom) Plot of $P/Q$ versus $x$ for pure TASEP (red points) and TASEP with reaction (blue points). For the latter, $P/Q>1$ and grows with  $x$. This suggests an effective death rate larger than $\lambda$ that grows with $x$, which is consistent with the top figure.}
 \label{fig-6}
\end{figure}

Across all the figures Fig.~\ref{fig-1}-Fig.~\ref{fig-6}, our MCS studies reveal that  $P/Q$ in general is space-dependent and larger than 1 in {\em all} the phases admitted by our model. This is in contrast to the pure TASEP limit, for which $P/Q\approx 1$ {\em always}. This shows that the effective decay rate is space-dependent and is bigger than its mean-field value. This in turn explains why MFT over-estimates the stationary densities and is always larger than the corresponding MCS result. Indeed, in the regions where $(P/Q -1)$ is bigger, the disagreement between MFT and MCS results are also bigger. This establishes that neglecting the correlation effects lies at the heart of the quantitative inaccuracy of MFT.

Our representative MCS studies on the staionary densities reveal LD, HD and LD-HD two phase coexistences, i.e., domain walls. Our limited studies have not detected the MC phase or mixed phase densities involving the MC phase. Detection of these phases require additional extensive parameter scnas, which are beyond the scope of the present study.

%[Plots of A/B versus x, MFT and MCS density profiles in the different MFT phases]

%Pure TASEP: A/B=1.

%\section{Results from improved MFT}

\section{Summary and outlook}\label{summ}

In this work, we have studied the steady states of an open totally asymmetric simple exclusion process (TASEP) coupled to bulk birth and death processes. The reactions are configuration-dependent: particle production requires an occupied site adjacent to a vacancy, whereas particle annihilation occurs between neighbouring occupied sites. This coupling of directed transport, exclusion and local reactions provides a minimal setting in which to investigate the interplay between nonequilibrium transport and density-dependent particle production and annihilation.

We have constructed a mean-field theory (MFT) for this model and analysed its stationary solutions in the hydrodynamic limit. For equal birth and death rates, the theory admits spatially varying density profiles in addition to a uniform solution with density 1/2. The resulting mean-field phase diagram contains low-density and high-density regimes, a maximal-current regime, and mixed phases in which distinct stationary density profiles occupy different portions of the system. The phase structure resembles that of TASEP with Langmuir kinetics, despite the different microscopic origins of the particle-nonconserving processes. In the present model,  the spatial variation of the stationary density is nonlinear in the continuum coordinate and is determined by the coupled transport-reaction balance.

Our Monte Carlo simulations reveal a substantial quantitative discrepancy between the mean-field stationary density profiles and those obtained from the stochastic dynamics. In the parameter regimes examined, MFT generally predicts higher densities than the simulations. This discrepancy is noteworthy because mean-field descriptions often provide useful quantitative predictions for open driven systems with short-ranged density correlations. The present results indicate that the microscopic structure of the reaction processes can be important even when the corresponding mean-field phase structure appears conventional.

We have investigated the origin of this discrepancy by examining nearest-neighbour occupation correlations. The exact evolution equation for the mean density depends on the probability of finding two neighbouring occupied sites, whereas MFT replaces this pair probability by the product of the corresponding one-site densities. This factorisation neglects correlations generated by the reaction dynamics. In particular, the birth process requires occupied-vacant pairs, while the death process acts on occupied-occupied pairs. Consequently, correlations between neighbouring sites modify the relative frequencies of production and annihilation events and can shift the stationary density away from the mean-field prediction.

Our numerical measurements of the nearest-neighbour pair probability provide evidence that this correlation effect is important in the present model. The results are consistent with the physical picture that the formation of occupied clusters suppresses particle production relative to the factorised estimate and enhances the contribution of pair annihilation. These effects can qualitatively account for the observed tendency of MFT to overestimate the stationary density. A complete quantitative description, however, requires the correlation corrections to be incorporated into the transport-reaction balance, including their spatial dependence. %In particular, the sign and magnitude of the correction must be examined across the different stationary regimes rather than inferred from a single global trend.

The comparison with TASEP with Langmuir kinetics highlights an important distinction between microscopic reaction mechanisms and their mean-field representations. Models with similar mean-field equations can exhibit different quantitative behaviour when their reactions depend differently on local particle configurations. In the present model, the correlations generated by the reactions  can substantially affect the predictions of the mean-field theory. This observation motivates a broader investigation of the conditions under which mean-field descriptions remain reliable in driven systems with particle-nonconserving dynamics.

Several questions remain open, which can form potential future research directions. First, it would be useful to explore the finite-size scaling of the discrepancy between MFT and Monte Carlo results and to determine whether it persists in the thermodynamic limit. Second, a systematic study of the connected nearest-neighbour correlation function across the different stationary regimes could clarify the spatial structure of the corrections to correlations and their relation to the density-profile errors. Third, extending the mean-field theory to include pair correlations, or employing a suitable cluster approximation, may improve the quantitative predictions. Such approaches would also help to determine whether the observed discrepancy can be understood predominantly through nearest-neighbour correlations or whether considering higher-order correlations are necessary. For reasons of simplicity, our present study is limited to equal birth and death rates. How unequal birth and death rates may modify our results would be an important question to answer in the future. It would also be interesting to investigate whether the competition between directed transport and configuration-dependent reactions produces additional regimes when the birth and death rates are unequal. Lastly, developing a systematic theory of the correlation effects, and establishing their consequences for the stationary phases, remain important directions for future work.

 {\em Acknowledgement:-} The authors thank Prof. Abhik Basu for his valuable discussions. SM thanks ANRF (India) for partial financial
support through the ARG (MATRICS) programme (file no.: ANRF/ARGM/2025/000748/TS).

\end{document}